\documentclass[journal]{IEEEtran}

\usepackage{hyperref}
\hypersetup{hidelinks}
\usepackage{xcolor,soul,framed} 

\colorlet{shadecolor}{yellow}
\usepackage[pdftex]{graphicx}
\DeclareGraphicsExtensions{.pdf,.jpeg,.png}

\usepackage{amsfonts,amssymb}
\usepackage{mathrsfs}
\usepackage{amsthm}
\usepackage[cmex10]{amsmath}
\usepackage{mathtools}
\usepackage{bbm}
\DeclareMathAlphabet{\mathbbb}{U}{bbold}{m}{n}
\usepackage{array}
\usepackage{aligned-overset}
\usepackage{flushend}
\usepackage{physics}
\usepackage{cite}
\usepackage{makecell}
\usepackage{threeparttable}
\usepackage{comment}
\usepackage{colortbl}
\let\labelindent\relax
\usepackage{enumitem}

\theoremstyle{definition}

\theoremstyle{remark}
\newtheorem{remark}{Remark}

\newcommand{\pha}[1]    {\underline{#1}}

\begin{document}
\bstctlcite{IEEEexample:BSTcontrol}
    \title{Grid-Forming Control of Modular Dynamic Virtual Power Plants}
    \author{Xiuqiang~He,~\IEEEmembership{Member,~IEEE,}
    Josué~Duarte,
    Verena~Häberle,~\IEEEmembership{Graduate Student Member,~IEEE,}
    and~Florian~Dörfler,~\IEEEmembership{Senior Member,~IEEE} \vspace{-5mm}
  \thanks{This work was supported by the European Union’s Horizon Europe programme under agreement 101096197.}
  \thanks{X. He, V. Häberle, and F. Dörfler are with the Automatic Control Laboratory, ETH Zurich, 8092 Zurich, Switzerland. J. Duarte is with the department of electrical engineering, Escola Universitaria Salesiana de Sarria, 08017, Barcelona, Spain. Emails: xiuqhe@ethz.ch, jnduarte@euss.cat, verenhae@ethz.ch, dorfler@ethz.ch.}}

\maketitle


\begin{abstract}

This article explores a flexible and coordinated control design for an aggregation of heterogeneous distributed energy resources (DERs) in a dynamic virtual power plant (DVPP). The control design aims to provide a desired aggregate grid-forming (GFM) response based on the coordination of power contributions between different DERs. Compared to existing DVPP designs with an AC-coupled AC-output configuration, a more generic modular DVPP design is proposed in this article, which comprises four types of basic DVPP modules, involving AC- or DC-coupling and AC- or DC-output, adequately accommodating diverse DER integration setups, such as AC, DC, AC/DC hybrid microgrids and renewable power plants. The control design is first developed for the four basic modules by the aggregation of DERs and the disaggregation of the control objectives, and then extended to modular DVPPs through a systematic top-down approach. The control performance is comprehensively validated through simulation. The modular DVPP design offers scalable and standardizable advanced grid interfaces (AGIs) for building and operating AC/DC hybrid power grids.

\end{abstract}

\vspace{-3mm}

\section{Introduction}

\IEEEPARstart{G}{rid}-forming (GFM) control has emerged as a recognized approach in recent years for enhancing the stability and dynamic performance of modern power grids, particularly with the growing penetration of distributed energy resources (DERs). However, GFM control for DERs presents several challenges. First, DERs are integrated into various setups, including AC, DC, and AC/DC hybrid systems \cite{kumar2015control,amira2019ac,ansari2020comprehensive}, each necessitating different forms of energy conversion. Second, due to the diverse nature of primary energy sources, each type of DER often requires distinct control objectives depending on their power and energy capabilities across different time scales \cite{haberle2022control,haberle2024gridforming,pishbahar2023emerging}. Additionally, multiple types of DERs are typically integrated into hybrid configurations to leverage their complementary characteristics and enhance overall efficiency, such as hybrid energy storage systems (HESSs) \cite{babu2020comprehensiv}. Third, effective control strategies must satisfy specific dynamic response requirements like fast frequency response (FFR), as demanded by grid codes \cite{haberle2024gridcode}, ancillary service markets \cite{Fingrid2021technical}, or optimal specification criteria \cite{haberla2024optimal}. To address these challenges, the concept of \textit{advanced grid interfaces} (AGIs) has been proposed by the EU's Horizon project AGISTIN \cite{agistin2024}. AGIs play a key role in controlling energy transfer between DERs and the grid via power converters, coordinating the dynamic response of heterogeneous DERs, and achieving the desired overall dynamic performance. A similar but more familiar concept is \textit{dynamic virtual power plants} (DVPPs), proposed by the earlier Horizon project POSYTYF \cite{posytyf2021}, which has gained broader attention \cite{ochoa2023control,zhu2024dynamic,hemin2024enhanced}. This article focuses on exploring GFM control designs for AGIs and DVPPs.

\begin{figure*}
  \begin{center}
  \includegraphics{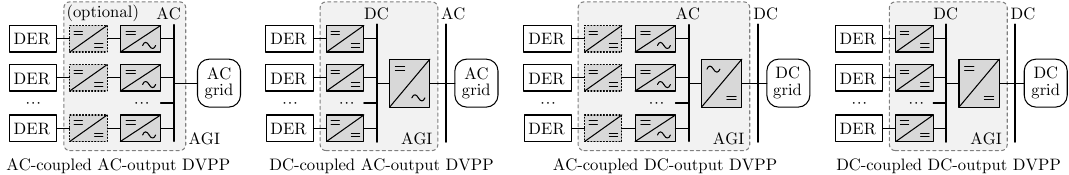}
  \caption{Four basic DVPP configurations, where DVPPs consist of AGIs and DERs. The AC-coupled AC-output configuration can be a renewable power plant or AC microgrid; the DC-coupled AC-output one can be a grid-connected DC microgrid; the AC-coupled DC-output one can be a high-voltage DC-integrated renewable power plant; the DC-coupled DC-output one can be a DC subgrid such as a DC data center or a DC charging station within a DC microgrid.}
  \label{fig:basic-dvpp}\vspace{-5mm}
  \end{center}
\end{figure*}

Several basic DVPP configurations are illustrated in Fig.~\ref{fig:basic-dvpp}, where DVPPs consist of AGIs and DERs. Precisely, AGIs are generic grid interfaces that collectively integrate, coordinate, and aggregate DERs (involving generation, storage, loads, etc.) to provide desired dynamic responses to the grid. The AGI configurations shown in Figs.~\ref{fig:basic-dvpp} and \ref{fig:modular-dvpp} include various AC and DC coupling types, as well as AC and DC output types, adequately representing the diverse topological arrangements for integrating DERs into power grids \cite{kumar2015control,amira2019ac,ansari2020comprehensive}, such as AC, DC, AC/DC hybrid microgrids, hybrid power plants, aggregation of DERs. Topologically, both emerging DVPPs and conventional microgrids aim to integrate and operate multiple DERs. However, from a control objective point of view, DVPPs are distinguished by their focus on the collective response to fulfill a desired response to the output-terminal grid, as opposed to the localized operation of individual DERs in a microgrid. A survey for localized power control and energy management in DC, AC, or AC/DC hybrid microgrids can be found in \cite{kumar2015control}, \cite{amira2019ac}, and \cite{ansari2020comprehensive}, respectively. To further enable a microgrid to provide GFM behaviors/services to the output-terminal grid, GFM controls such as droop control \cite{gu2015frequency,macana2022distributed}, virtual synchronous machine control \cite{mohamed2022optimal,sun2022design}, matching control \cite{chen2017integration}, or dual-port power-balancing control \cite{subotic2022dualport} have been applied in the grid-connected inverter of microgrids.

Despite considerable research on microgrid control, aggregation of DERs, and provision of GFM behaviors, existing microgrid control designs struggle to meet the flexible control objectives required by grid codes and enabled by DVPPs. Particularly, there are two significant gaps in existing control approaches. First, existing approaches may lack the flexibility to deliver a desired aggregate response (typically dictated by grid code requirements \cite{haberle2024gridcode} or an optimal specification \cite{haberla2024optimal}) to the output-terminal grid. Second, existing control designs are specifically tailored, while not easily scalable to accommodate AC/DC hybrid systems. In our recent work \cite{haberle2022control,haberle2024gridforming}, a flexible DVPP control design is advocated. However, this DVPP design is limited to an AC-coupled configuration and the output terminal can only be connected to an AC grid. For this reason, there is a lack of more generic DVPP designs that address AC, DC, and AC/DC hybrid configurations.

Therefore, this study explores a wide range of DVPP configurations including AC-coupled, DC-coupled, AC-output, and DC-output configurations, i.e., presenting four basic DVPP modules. These basic modules can be interconnected to form a modular AC/DC hybrid DVPP. A modular DVPP design offers high flexibility in achieving a desired dynamic response, high scalability in modular configuration and control, uniform standardization for multi-AC/DC-port interconnections, and high transparency of dynamic responses. The major contributions of this article are summarized as follows.
\begin{itemize}
    \item We present four basic DVPP configurations/modules that represent four typical topological arrangements in the current DER integration practice. For each of these, we formulate desired and interoperable control objectives.
    \item We propose a control design for each basic DVPP module, including the condition for DER aggregation, the coordinated disaggregation of the desired behavior, and the local control strategies for individual converters.
    \item We extend this control design to encompass modular AC/DC hybrid DVPPs through a systematic top-down design approach, offering a modular DVPP solution with high flexibility and scalability. 
\end{itemize}

The remainder of this article is organized as follows. Section~\ref{sec:dvpp-topology-objectives} introduces basic and modular DVPP configurations and the associated control objectives. The control design is detailed in Section~\ref{sec:control-design}. Simulation validations are presented in Section~\ref{sec:validations}. Section~\ref{sec:conclusion} concludes this article.

\section{DVPP Configurations and Control Objectives}
\label{sec:dvpp-topology-objectives}

This section first presents four basic DVPP modules, then shows a modular DVPP configuration, and lastly introduces their control objectives.

\subsection{Basic DVPP Modules}

We show four basic DVPP configurations in Fig.~\ref{fig:basic-dvpp} with different coupling and terminal output types:
\begin{itemize}
    \item AC-coupled AC-output topology, with separate DC-AC converters (and without a central converter);
    \item DC-coupled AC-output topology, with separate DC-DC converters and a central DC-AC converter;
    \item AC-coupled DC-output topology, with separate DC-AC converters and a central AC-DC converter;
    \item DC-coupled DC-output topology, with separate DC-DC converters and a central DC-DC converter.
\end{itemize}
Among these topologies, only the AC-coupled AC-output DVPP configuration has been explored \cite{haberle2022control,haberle2024gridforming}. The other topologies have not been studied in the context of DVPPs, albeit similar topological arrangements are widely present in AC/DC microgrids or renewable power plants \cite{kumar2015control,amira2019ac,ansari2020comprehensive}. Recognizing the existence of all these topologies, we aim to explore generic modular configurations and control designs.

\begin{figure}
  \begin{center}
  \includegraphics{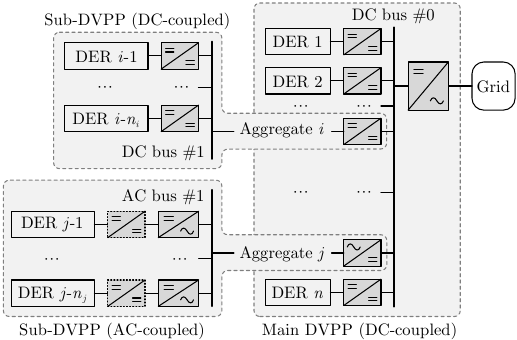}
  \caption{A modular AC/DC hybrid DVPP, where a DC-coupled AC-output module (main DVPP), a DC-coupled DC-output module (sub-DVPP), and an AC-coupled DC-output module (sub-DVPP) are included.}
  \label{fig:modular-dvpp}\vspace{-4mm}
  \end{center}
\end{figure}

\subsection{Modular DVPP Configuration}

Building on the basic modules, we can establish modular AC/DC hybrid DVPP configurations, an illustrative example of which is shown in Fig.~\ref{fig:modular-dvpp}. There are $n$ DERs in the top-level main DVPP module, which are connected at the DC bus \#0. The aggregate $i$ is represented by a collection of $n_i$ DERs connected at the DC bus \#1. The ensemble of these $n_i$ DERs forms a DC-coupled DC-output DVPP module (referred to as a sub-DVPP). Likewise, the aggregate $j$ is represented by a collection of $n_j$ DERs connected at the AC bus \#1, considered an AC-coupled DC-output sub-DVPP. A large AC/DC hybrid DVPP can be assembled in such a modular way by encompassing multiple DC and AC coupling buses.

We note that these basic and modular DVPP configurations are alternatively known in the literature as AC, DC, or AC/DC hybrid microgrids (or hybrid power plants) \cite{kumar2015control,amira2019ac,ansari2020comprehensive}. However, we refer to them as ``DVPPs" in this study from the output-terminal control perspective. In other words, we propose to control the DVPPs in an aggregated way to collectively provide desired dynamic responses, particularly, GFM behaviors, to the output-terminal AC or DC grid. In contrast, conventional microgrid controls mainly focus on power and energy management \textit{within} the microgrid \cite{kumar2015control,amira2019ac,ansari2020comprehensive}. Therefore, despite having similar topological arrangements, DVPPs have distinct operation objectives compared to conventional microgrids. More importantly, a modular DVPP introduces a flexible and scalable approach to building AC/DC hybrid power grids. This approach is more systematic and standardized than the conventional fragmented and self-contained integration paradigm of DERs. Overall, a modular DVPP offers many advantages, including but not limited to the following.
\begin{itemize}
    \item High flexibility in achieving a desired dynamic behavior;
    \item High scalability in modular configuration and control;
    \item Uniform standardization in enabling standardized protocols for multi-AC/DC-port interconnections; 
    \item High transparency in understanding the dynamic behavior of any bus from the DVPP level to the DER level.
\end{itemize}

\subsection{Control Objectives of DVPP Modules}

\begin{figure}
  \begin{center}
  \includegraphics{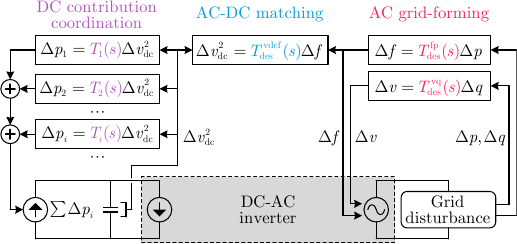}
  \caption{Control objective illustration of a DC-coupled AC-output DVPP.}
  \label{fig:control-objectives}\vspace{-3mm}
  \end{center}
\end{figure}

We consider a DC-coupled AC-output DVPP module (which can arise as grid-connected DC microgrids) and formulate its control objectives. The control objectives of the other modules can be formulated similarly and will be summarized at the end of this subsection. For the DC-coupled AC-output module under consideration, its control objectives are illustrated in Fig.~\ref{fig:control-objectives}, which comprise three parts: 1) the desired AC GFM response, 2) the AC-DC matching relationship, and 3) the DC contribution coordination. 

\subsubsection{AC GFM}

To provide a desired AC GFM dynamic behavior, we specify the AC frequency and voltage response of the central inverter under AC-side power disturbances as
\begin{equation}
    \label{eq:ac-gfm}
    \begin{bmatrix}
        \Delta f(s)\\ \Delta v(s)
    \end{bmatrix} = 
    \begin{bmatrix}
        T_\mathrm{des}^{\mathrm{fp}}(s) & 0 \\ 0 & T_\mathrm{des}^{\mathrm{vq}}(s) 
    \end{bmatrix}
    \begin{bmatrix}
        \Delta p(s) \\ \Delta q(s)
    \end{bmatrix},
\end{equation}
where $\Delta p$ and $\Delta q$ denote the active and reactive power output changes at the output terminal of the central inverter, and $\Delta f$ and $\Delta v$ denote the desired GFM response to the power disturbance. In the scope of dynamic ancillary services, we consider small-signal increment variables as the deviations of large-signal variables from their associated steady states or setpoints. The transfer functions in \eqref{eq:ac-gfm} can encode any desired linear/linearized responses in general such as virtual inertial response, damping, and steady-state droop relationship.  

\begin{remark}
The desired dynamic response in \eqref{eq:ac-gfm} is specified by grid codes \cite{oyj2023technical}. However, most current grid codes specify a desired dynamic response through piece-wise linear step-response capabilities curves in the time domain, rather than through a transfer function in the frequency domain. The latter, though, is preferable for control design and is more familiar to power electronics engineers. To obtain the transfer function corresponding to a dynamic response specification in a grid code, one can utilize the translation approach in \cite{haberle2024gridcode}. Therefore, we assume that the desired transfer functions in \eqref{eq:ac-gfm} are given, which satisfy grid code requirements and respect the maximum capacity limitations of the DVPP \cite{haberle2024gridcode}.
\end{remark}

\begin{table*}
\centering
\caption{Control Objectives and Control Designs of the Four Basic DVPP Modules} \vspace{-3mm}
\begin{tabular}[t]{c|c|c|c|c}
\hline \hline
DVPP module & AC-coupled AC-output  & DC-coupled AC-output & AC-coupled DC-output & DC-coupled DC-output \\
\hline
\makecell{Coupling type}& AC & DC & AC & DC \\ 
\makecell{Output type} & AC & AC & DC & DC  \\ 
\makecell{Desired output behavior} &
$\Delta f = T_\mathrm{des}^{\mathrm{fp}} \Delta p$ &
$\Delta f = T_\mathrm{des}^{\mathrm{fp}} \Delta p$ &
$\Delta p = T^{\mathrm{pvdc}}_{\mathrm{des}} \Delta (v_{{\mathrm{dc}}}^2)$ & 
$\Delta p = T^{\mathrm{pvdc}}_{\mathrm{des}} \Delta (v_{{\mathrm{dc}}}^2)$ \\
\makecell{Desired matching between \\ the coupling and output buses} & 
$\Delta {f} = \Delta {f}$ & 
$\Delta (v_{{\mathrm{dc}}}^2) = T_{{\mathrm{des}}}^{{\mathrm{vdcf}}} \Delta {f}$ & 
$\Delta (v_{{\mathrm{dc}}}^2) = T_{{\mathrm{des}}}^{{\mathrm{vdcf}}} \Delta {f}$ & 
$\Delta (v_{{\mathrm{dc}},1}^2) = T_{{\mathrm{des}}}^{{\mathrm{vdcvdc}}} \Delta (v_{{\mathrm{dc}}}^2)$ \\
\makecell{Individual DER contribution} & 
$\Delta f_i = T_i \Delta p_i$ & 
$\Delta p_i = T_i \Delta (v_{{\mathrm{dc}}}^2)$ & 
$\Delta f_i = T_i \Delta p_i$ & 
$\Delta p_i = T_{i} \Delta (v_{{\mathrm{dc}},1}^2)$ \\
\makecell{Aggregation condition} & 
$\sum\nolimits_i T_i^{-1} = T_\mathrm{des}^{\mathrm{fp}-1}$ & 
$\sum\nolimits_{i} T_i = {T_{{\mathrm{des}}}^{{\mathrm{vdcf}-1}} T_\mathrm{des}^{\mathrm{fp}-1} }$ & 
$\sum\nolimits_i T_i^{-1} =  T_\mathrm{des}^{\mathrm{vdcf}} T_{{\mathrm{des}}}^{{\mathrm{pvdc}}}$ & 
$\sum\nolimits_i T_i = T_{{\mathrm{des}}}^{{\mathrm{vdcvdc}}{-1}} T^{\mathrm{pvdc}}_{\mathrm{des}}$ \\
Reference & \cite{haberle2024gridforming}; Appendix & Section III & Appendix & Appendix \\
\hline \hline
\end{tabular}\vspace{-3mm}
\label{tab:summary-dvpp-modules}
\end{table*}

\subsubsection{AC-DC Matching}

A DC-coupled DVPP must maintain a stable and properly bounded DC-bus voltage to ensure normal operation. Inspired by the well-known matching control \cite{huang2017virtual,arghir2018grid,subotic2022power}, to perform a DC-bus voltage forming behavior, we specify an AC-DC matching relationship as,
\begin{equation}
    \label{eq:ac-dc-matching}
    \Delta (v_{{\mathrm{dc}}}^2)(s) = T_{{\mathrm{des}}}^{{\mathrm{vdcf}}}(s) \Delta {f}(s),
\end{equation}
where $\Delta (v_{{\mathrm{dc}}}^2)$ denotes the variation of the DC voltage square\footnote{The reason for using voltage square, as in \cite{huang2017virtual}, is because power is linear in the square of voltage, simplifying the control design.} in response to the AC-side disturbance. The AC-DC matching in \eqref{eq:ac-dc-matching} thus specifies a desired relationship between the AC and DC ports of the inverter. The relationship not only facilitates the control design as shown in the next section but also offers a straightforward way to form the DC voltage and constrain its variation within an allowed range. More specifically, given an allowed variation range of the grid frequency, $\vert \Delta f \vert_{\max}$, we can specify an appropriate $T_{{\mathrm{des}}}^{{\mathrm{vdcf}}}(s)$ such that the response of $\vert \Delta (v_{{\mathrm{dc}}}^2) \vert$ is within an allowed variation range $\vert \Delta (v_{{\mathrm{dc}}}^2) \vert_{\max}$.

\subsubsection{DC Contribution Coordination}

The desired GFM response ultimately relies on the power provision of DERs. To achieve the desired responses in \eqref{eq:ac-gfm} and \eqref{eq:ac-dc-matching}, the control of DERs must be coordinated appropriately since heterogeneous DERs have different power response characteristics over time scales. In the frequency domain, the power response of each DER to a DC voltage variation is specified as
\begin{equation}
\label{eq:DER-dynamics}
    \Delta p_i(s) = T_i(s) \Delta (v_{{\mathrm{dc}}}^2)(s),
\end{equation}
where $T_i(s)$ represents the closed-loop control behavior within each DER and $\Delta p_i(s)$ denotes the associated power output change. The specification of the control behavior $T_i(s)$ should consider the following factors:
\begin{itemize}
    \item The specification of $T_i(s)$ must respect the DER's characteristics. For example, a supercapacitor can only provide a dynamic response but not a steady-state contribution.
    \item For non-controllable DERs or user loads, their dynamics, captured by $T_i(s)$, are assumed to be fixed and known.
    \item The collective response of $T_i(s)$ should fulfill the desired AC GFM response in \eqref{eq:ac-gfm} and the AC-DC matching relationship in \eqref{eq:ac-dc-matching}.
\end{itemize}

Analogously, we can specify the control objectives of the other DVPP modules, which are summarized in Table~\ref{tab:summary-dvpp-modules} and outlined in Appendix. We highlight that the control objectives are interoperable between different DVPP modules, thus allowing the interconnection of multiple modules.

Finally, to satisfy the control objectives of a DVPP module, the control behaviors of all converters in the module should be coordinated. A coordinated GFM control design is presented in the next section to achieve the control objectives.

\section{Control Design of DVPPs}
\label{sec:control-design}

This section presents a control design for a basic DVPP module and then extends it to a modular DVPP.

\subsection{DER Aggregation Condition}

To disaggregate the desired control objectives in \eqref{eq:ac-gfm} and \eqref{eq:ac-dc-matching} into the individual DER control behaviors in \eqref{eq:DER-dynamics}, we need first to aggregate the DER control behaviors. Based on \eqref{eq:DER-dynamics}, the aggregated power response on the DC bus is
\begin{equation}
\label{eq:aggregated-dc-dynamics}
    \sum\nolimits_{i = 1}^n \Delta p_i(s) = \sum\nolimits_{i = 1}^n T_i(s) \Delta (v_{{\mathrm{dc}}}^2)(s),
\end{equation}
where there are $n$ devices, including the controllable DERs (with DC-DC converters), non-controllable devices, and the DC capacitor (note that the capacitor also participates in the power response). For example, the dynamics of the DC capacitor $C_{\mathrm{dc}}$ is given as $T_{\mathrm{c}}(s) = -\frac{1}{2}C_{\mathrm{dc}} s$, with the negative sign complying with the sign convention of the output power. The power conservation between the DC and AC ports of the central inverter is represented as
\begin{equation}
    \label{eq:power-conservation}
    \sum\nolimits_{i = 1}^n \Delta p_i(s) = \Delta p(s),
\end{equation}
where we neglect the power losses in the DC network and converters. By plugging \eqref{eq:aggregated-dc-dynamics} and \eqref{eq:power-conservation} into \eqref{eq:ac-gfm} and \eqref{eq:ac-dc-matching}, we derive the following aggregation condition
\begin{equation}
    \label{eq:aggregation}
    \sum\nolimits_{i = 1}^n T_i(s) = T_{{\mathrm{des}}}^{{\mathrm{vdcf}}}(s)^{-1} T_\mathrm{des}^{\mathrm{fp}}(s)^{-1}.
\end{equation}
It indicates the relationship in which the aggregated dynamics of the DC-side devices should satisfy the desired response.

For non-controllable devices, we account for their inherent dynamics as $T_i(s)$, e.g., the DC capacitor dynamics $T_{\mathrm{c}}(s)$ shown earlier. We denote by $\mathcal{N}$ the set of non-controllable devices and by $\mathcal{C}$ the set of controllable DERs. Thus, the aggregation condition in \eqref{eq:aggregation} becomes
\begin{equation}
    \label{eq:aggregation1}
    \sum\nolimits_{i \in \mathcal{C}} T_i(s) + \sum\nolimits_{i \in \mathcal{N}} T_i(s) = T_{{\mathrm{des}}}^{{\mathrm{vdcf}}}(s)^{-1} T_\mathrm{des}^{\mathrm{fp}}(s)^{-1}.
\end{equation}

\begin{figure}
  \begin{center} \includegraphics{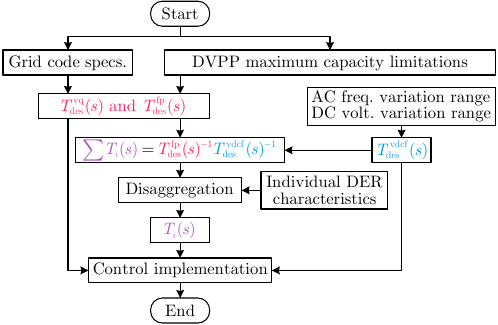}
  \caption{Flowchart of the control design of a DC-coupled AC-output DVPP.}
  \label{fig:flowchart}\vspace{-3mm}
  \end{center}
\end{figure}

\subsection{Coordinated Disaggregation of the Desired Behavior}

Based on the aggregation condition in \eqref{eq:aggregation1}, we disaggregate the desired response in the right-hand side of \eqref{eq:aggregation1} appropriately, obtaining $T_i(s)$ for each controllable DER. To achieve this, we use dynamic participation factors (DPFs) similarly as for AC coupled-AC-output DVPPs \cite{haberle2022control}. Namely, we specify the contribution of each controllable DER, $T_i(s),\, {i \in \mathcal{C}}$, as
\begin{equation}
    \label{eq:contribution}
    T_i(s) \coloneqq m_i(s) \left(T_{{\mathrm{des}}}^{{\mathrm{vdcf}}}(s)^{-1} T_\mathrm{des}^{\mathrm{fp}}(s)^{-1} - \sum\nolimits_{i \in \mathcal{N}} T_i(s) \right).
\end{equation}
The condition in \eqref{eq:aggregation1} then simplifies to $    \sum\nolimits_{i \in \mathcal{C}} m_i(s) = 1$. For fast-acting DERs without steady-state contribution, e.g., supercapacitors or flywheels, we can choose a high-pass filter for $m_i(s)$ to provide a fast dynamic response. For relatively slow DERs with steady-state contribution, e.g., renewable energy reserve units, we can choose a low-pass filter for $m_i(s)$ to provide power support on a longer time scale. For DERs covering the intermediate regime, e.g., battery energy storage systems, we can specify $m_i(s)$ as the residual part to fulfill the aggregation condition. The dynamic and steady-state participation can be adapted by tuning the time constant and static gain in $m_i(s)$. As in \cite{haberle2022control} for AC-coupled AC-output DVPPs, the adaption of the DPFs can be performed online in a centralized manner to account for the plugging-in and plugging-out of DERs and time-varying capacity limitations depending on weather conditions.

\begin{figure*}
  \begin{center}\vspace{-4mm}
  \includegraphics{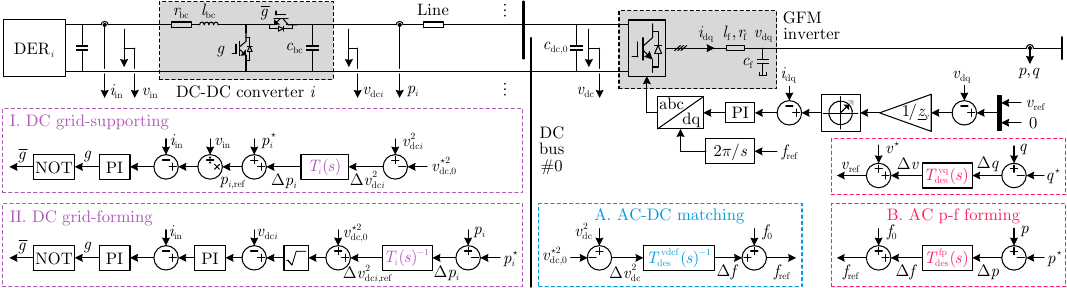}
  \caption{Control block diagram of a DC-coupled AC-output DVPP, displaying two control implementation options for the inverter: A. AC-DC matching and B. AC p-f forming, and two control implementation options for the DC-DC converters: I. DC grid-supporting and II. DC grid-forming. Note that, the aggregation of $T_i(s)$ satisfies the condition in \eqref{eq:aggregation} concerning both $T_{{\mathrm{des}}}^{{\mathrm{vdcf}}}(s)$ and $ T_\mathrm{des}^{\mathrm{fp}}(s)$ and therefore, the desired behavior for both can be achieved, even though only one (i.e., option A or B) is implemented explicitly in the inverter, as the other one is achieved implicitly.}
  \label{fig:control-block}\vspace{-5mm}
  \end{center}
\end{figure*}

\subsection{Implementable Local Control Strategies}

The flowchart of the above control design is presented in Fig.~\ref{fig:flowchart}, where the procedure for obtaining desired control behaviors from the grid level to the local device level is shown. The obtained transfer function $T_{{\mathrm{des}}}^{{\mathrm{vdcf}}}(s)$ or $T_\mathrm{des}^{\mathrm{fp}}(s)$ will be used as frequency control specification of the central inverter, $T_\mathrm{des}^{\mathrm{vq}}(s)$ as voltage control specification of the inverter, and $T_i(s)$ as control specification of the DC-DC converters. An implementable control block diagram is shown in Fig.~\ref{fig:control-block}.

\subsubsection{Inverter Control}
The frequency control of the inverter can employ either $T_{{\mathrm{des}}}^{{\mathrm{vdcf}}}(s)$ or $T_\mathrm{des}^{\mathrm{fp}}(s)$ as the outer-loop controller, as shown in Fig.~\ref{fig:control-block}, allowing two different implementations: A. AC-DC matching and B. AC p-f forming. The former uses the DC voltage while the latter uses the AC power as the feedback signal. More generally, one can also use their linear combination, which resembles a dual-port GFM control \cite{subotic2022dualport}. We recommend the AC-DC matching one between these two implementations since it does not require power setpoint coordination with the DC-DC converters. It also shows more robust DC voltage stability under large grid disturbances as it can self-balance the DC-bus power between the DER injection and the inverter output \cite{tayyebi2020frequency}.

Compared to the frequency control, the voltage control of the inverter is simple (without need for DER coordination), which directly takes $T_\mathrm{des}^{\mathrm{vq}}(s)$ as the outer-loop controller, closing the loop between reactive power and voltage magnitude.

Given the resulting frequency reference $f_{\mathrm{ref}}$ and voltage magnitude reference $v_{\mathrm{ref}}$ from the outer-loop control outputs, the inner voltage and current loops can use proportional-integral (PI) controllers. One can replace the voltage PI controller with a virtual admittance for adaptive current limiting and fault ride-through under grid faults. The transient stability of the inverter can be further enhanced with an additional feedback loop of current saturation information, which can be found in \cite{he2024cross}, and is not included in Fig.~\ref{fig:control-block}.

\subsubsection{DC-DC Converter Control}

The control of the DC-DC converters also admits multiple implementations, two of which are shown in Fig.~\ref{fig:control-block}: I. DC grid-supporting and II. DC grid-forming. The former, similar to classical AC grid-supporting control, measures the voltage variation and adjusts the power reference. The latter, resembling classical AC GFM control, measures the output power change and regulates the voltage reference. The desired control behavior $T_i(s)$ and its inverse $T_i(s)^{-1}$ are used as the outer-loop controllers in the two implementations, respectively. By comparison, we recommend the DC grid-supporting implementation since power limitations can be easily implemented by incorporating a limiter to the power reference $p_{i,\mathrm{ref}}$. In contrast, power limitations are challenging for the DC grid-forming implementation. Similar to the inverter, the inner voltage and current loops of the DC-DC converter can use PI controllers. 

\begin{remark}
All transfer-function controllers to be implemented must be causal and stable. Since the desired DVPP response is typically stable, the disaggregated transfer functions are also stable. Furthermore, if one transfer function is non-causal, we can causalize it by adding a stable filter with a small time constant. We further note that PI controllers for the converter inner loops are basic solutions, which can be replaced by more advanced controllers such as $\mathcal{H}_{\infty}$ control \cite{haberle2022control}.
\end{remark}

\subsection{Extension to Modular AC/DC Hybrid DVPPs}

The control design for the four basic DVPP modules differs from each other due to their different coupling and output types. To be specific, the main difference lies in their aggregation conditions while the disaggregation and implementation are similar. For the DC-coupled AC-output DVPP, the aggregation condition is already given in \eqref{eq:aggregation}. We derive the aggregation conditions for the other three modules in Appendix and summarize their control designs in Table~\ref{tab:summary-dvpp-modules}.

As with the modular DVPP configuration, we now extend the control design to modular DVPPs. Since a modular DVPP can be assembled module by module, its control design can be developed in a systematic top-down approach, given the aggregation conditions for the individual DVPP modules. Consider the AC/DC hybrid DVPP shown earlier in Fig.~\ref{fig:modular-dvpp} as an example. The top-down control design is outlined as follows.
\begin{itemize}
    \item At the grid level, the desired AC GFM behavior is specified as in \eqref{eq:ac-gfm}.
    \item At the main-DVPP level, the control design for the DC-coupled AC-output module is based on the aggregation condition in \eqref{eq:aggregation} and the disaggregation in \eqref{eq:contribution}.
    \item At one sub-DVPP level, the control design for the DC-coupled DC-output module is based on the aggregation condition in \eqref{eq:dc-dc-aggregation-condition} in Appendix, which inherits the aggregate $i$'s desired control behavior $T_i(s)$ as the desired output behavior of this sub-DVPP module.
    \item At another sub-DVPP level, the control design for the AC-coupled DC-output module is based on the aggregation condition in \eqref{eq:ac-dc-aggregation-condition} in Appendix, which inherits the aggregate $j$'s desired control behavior $T_j(s)$ as the desired output behavior of this sub-DVPP module.
    \item Following the same way until we reach the bottom level, obtaining the desired control behavior of each converter.
\end{itemize}

\begin{figure}
  \begin{center}
  \includegraphics{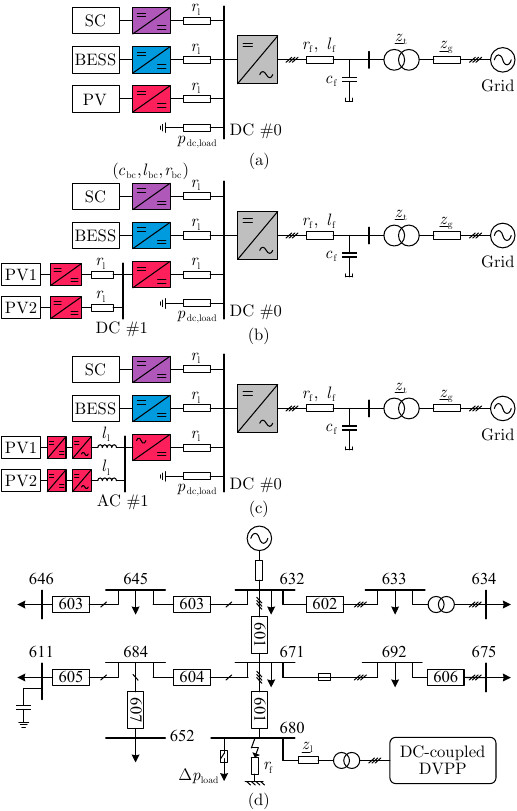}\vspace{-3mm}
  \caption{Illustration of the system setup in case studies. (a) A basic DC-coupled AC-output DVPP. (b) A multi-DC-bus DVPP. (c) An AC/DC hybrid DVPP. (d) A DC-coupled AC-output DVPP connected to the IEEE 13-bus system.}
  \label{fig:case-study-system}\vspace{-8mm}
  \end{center}
\end{figure}

\begin{figure}[t]
  \begin{center}
  \includegraphics{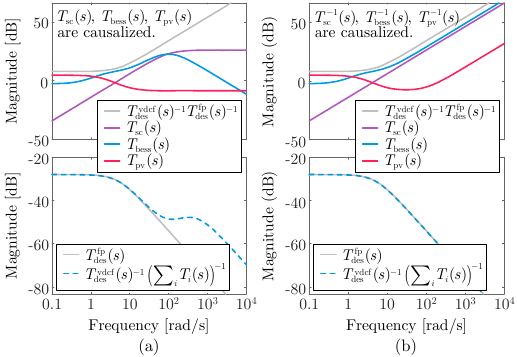}
  \caption{Magnitude Bode plots, where $T_{{\mathrm{des}}}^{{\mathrm{vdcf}}-1} T_\mathrm{des}^{\mathrm{fp}-1}$ denotes the desired power response; $T_{\mathrm{sc}}$, $T_{\mathrm{bess}}$, and $T_{\mathrm{pv}}$ the power responses of the DERs; $T_{\mathrm{des}}^{\mathrm{fp}}$ the desired frequency response; and $T_{{\mathrm{des}}}^{{\mathrm{vdcf}}-1} \left(\sum\nolimits_i T_i\right)^{-1}$ the desired frequency response using causalized $T_i$ or $T_i^{-1}$. (a) For the DC grid-supporting implementation. (b) For the DC grid-forming implementation.}
  \label{fig:case-study-A1-bode}\vspace{-5mm}
  \end{center}
\end{figure}

\begin{remark}
The top-down control design focuses on the AC-bus frequency and DC-bus voltage responses, requiring coordination among all AC and DC buses across converters. In contrast, the AC voltage response of AC buses can be controlled locally, without requiring coordination with other AC and DC buses. For each AC bus in a modular DVPP, the voltage control for a desired response can be addressed using the same design as for an AC-coupled AC-output DVPP in \cite{haberle2022control,haberle2024gridforming}.
\end{remark}

\section{Simulation Validations}
\label{sec:validations}

\begin{figure*}
  \begin{center}
  \includegraphics{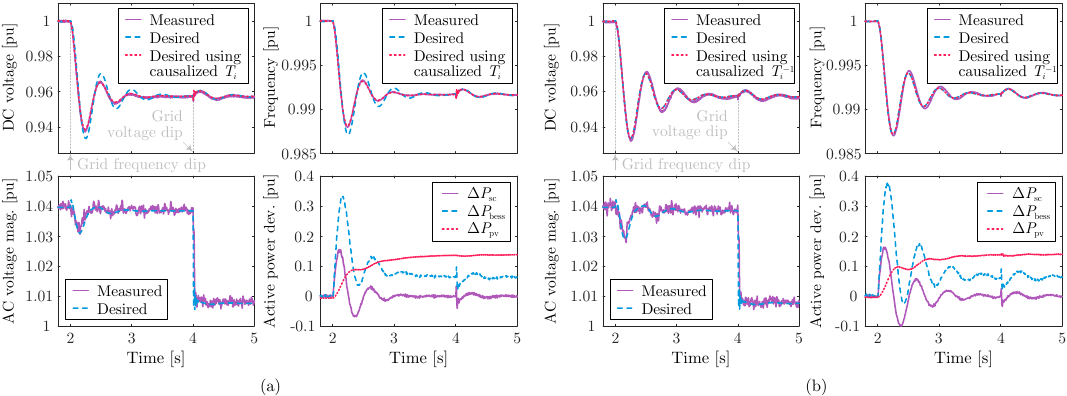}\vspace{-3mm}
  \caption{Simulation results of case study A1: the basic DVPP in Fig.~\ref{fig:case-study-system}(a) under a grid frequency dip and a voltage dip. The control of the DC-DC converters uses (a) DC grid-supporting implementation and (b) DC grid-forming implementation. The ripples in the waveforms are due to the IGBT switching.}
  \label{fig:case-study-A1}\vspace{-1mm}
  \end{center}
\end{figure*}

\begin{figure*}
  \begin{center}
  \includegraphics{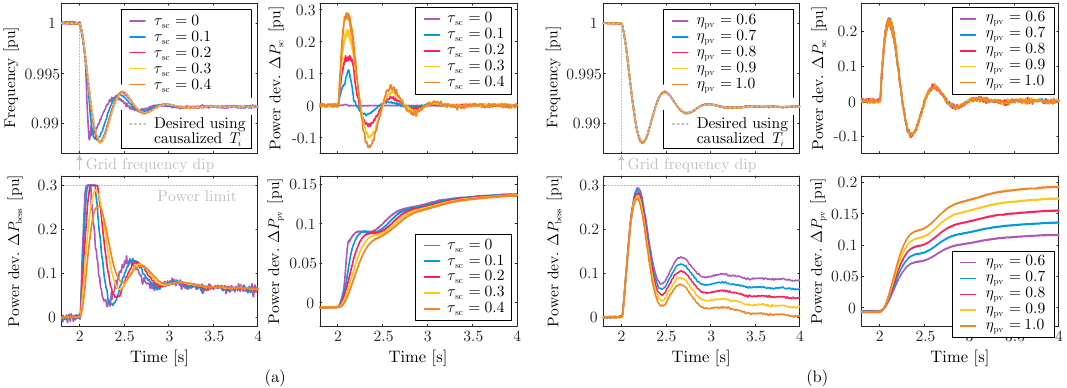}\vspace{-3mm}
  \caption{Simulation results of case study A2: the basic DVPP in Fig.~\ref{fig:case-study-system}(a) with different coordination between the DERs. (a) Different inertia time constants $\tau_{\mathrm{sc}}$ in the SC's power response $T_{\mathrm{sc}} = -\frac{\tau_{\mathrm{sc}} s}{0.01s + 1}$. (b) Different participation factor static gains $\eta_{\mathrm{pv}}$ in the PV's power response $T_{\mathrm{pv}} = \tfrac{\eta_{\mathrm{pv}}}{0.5s + 1} T_{\mathrm{bess\&pv}} $.}
  \label{fig:case-study-A2}\vspace{-3mm}
  \end{center}
\end{figure*}

This section presents case studies to validate the performance of the proposed control design. The simulation system setup is shown in Fig.~\ref{fig:case-study-system}. In case study~A, a DVPP is connected to an infinite-bus system to test whether it can achieve the desired response. The infinite-bus setup is used to impose an intended change in grid frequency and voltage that complies with the test requirements of grid codes \cite{oyj2023technical}. In case study~B, we connect the DVPP to the IEEE 13-bus distribution system \cite{ieee13node} to test its performance under more realistic load disturbances and short-circuit faults. The case studies employ high-fidelity simulation models of DC-AC and DC-DC converters, in which their inner voltage and current loops, PWM modulation, and IGBT switches are included. The system parameters are provided in Table~\ref{tab:system-parameters} in Appendix.

The desired dynamic response is represented in per-unit as
\begin{equation}
\label{eq:bess-pv}
    T_{\mathrm{des}}^{\mathrm{fp}} = \frac{-1}{5s + 25}, \quad T_{\mathrm{des}}^{\mathrm{vq}} = \frac{-0.2}{0.01s + 1}, \quad T_{\mathrm{des}}^{\mathrm{vdcf}} = 10,
\end{equation}
where we omit ``$(s)$" for brevity, similarly hereafter, with minus signs indicating negative feedback droop, an inertia time constant $5$\,s, a typical frequency droop sloop $1/25 = 4$\%, and a typical voltage droop sloop $20$\%. Moreover, the desired AC-DC matching relationship $T_{\mathrm{des}}^{\mathrm{vdcf}}$ suggests that a $10$\% DC voltage-square variation will accompany a $1$\% frequency variation. The DVPP comprises a supercapacitor (SC), a better energy storage system (BESS), and a photovoltaic (PV) system (power-reserved), with power setpoints as $0.0$, $0.0$, and $0.5$\,pu, respectively. To disaggregate the desired power response, $T_{\mathrm{des}}^{\mathrm{vdcf}-1} T_{\mathrm{des}}^{\mathrm{fp}-1}$, we pre-specify the SC's response as $T_{\mathrm{sc}} \coloneqq -\tau_{\mathrm{sc}} s$, $\tau_{\mathrm{sc}} \geq 0$ [one can alternatively assign a high-pass filter $m_i(s)$ for it as in \eqref{eq:contribution}], considering its fast-acting capability to contribute an inertial response. Then, the desired collective response of the BESS and PV is represented as $T_{\mathrm{bess\&pv}} \coloneqq {T_{\mathrm{des}}^{\mathrm{vdcf}-1} T_{\mathrm{des}}^{\mathrm{fp}-1}} - T_{\mathrm{sc}} + \tfrac{1}{2}(c_{\mathrm{dc,0}} + 3c_{\mathrm{bc}})s$ using the aggregation condition in \eqref{eq:aggregation1}, where $\tfrac{1}{2}(c_{\mathrm{dc,0}} + 3c_{\mathrm{bc}})s$ denotes the non-controllable DC capacitors' response. Lastly, $T_{\mathrm{bess\&pv}}$ is disaggregated between the BESS and the PV using dynamic participation factors. We choose a low-pass filter $m_{\mathrm{pv}}(s) = \tfrac{\eta_{\mathrm{pv}}}{0.5s + 1}$ for the power-reserved PV to provide power support relatively slowly while for a longer time. The residual part $m_{\mathrm{bess}}(s) = \tfrac{0.5s+1 - \eta_{\mathrm{pv}}}{0.5s + 1}$ is assigned to the BESS. Their desired power responses are represented by $T_{\mathrm{pv}} = m_{\mathrm{pv}}(s) T_{\mathrm{bess\&pv}} $ and $T_{\mathrm{bess}} = m_{\mathrm{bess}}(s) T_{\mathrm{bess\&pv}}$.

\subsection{Case Study A: Connection to an Infinite-Bus System}

\subsubsection{Case Study A1: Grid Disturbances}

This study aims to validate the dynamic response of the basic DVPP in Fig.~\ref{fig:case-study-system}(a) under grid frequency and voltage dips. We test both control implementations shown in Fig.~\ref{fig:control-block} for the DC-DC converters: I. DC grid-supporting and II. DC grid-forming. For the former, we choose $T_{\mathrm{sc}} = -\frac{0.2 s}{0.01s + 1}$, $T_{\mathrm{pv}} = \tfrac{0.7}{0.5s + 1} T_{\mathrm{bess\&pv}}$, $T_{\mathrm{bess}} = \tfrac{0.5s+0.3}{0.5s + 1} T_{\mathrm{bess\&pv}}\frac{1}{(0.01s + 1)^2}$ (all in per-unit), where $T_{\mathrm{sc}}$ and $T_{\mathrm{bess}}$ are causalized by low-pass filters with a small time constant $0.01$. For the latter, we specify $T^{-1}_{\mathrm{sc}} = -\frac{1}{0.2s}$, $T^{-1}_{\mathrm{pv}} = \tfrac{0.5s + 1}{0.7} T^{-1}_{\mathrm{bess\&pv}} \frac{1}{0.01s + 1}$, and $T^{-1}_{\mathrm{bess}} = \frac{1}{T_{\mathrm{bess\&pv}} - T_{\mathrm{pv}}}$. The Bode diagrams of these transfer functions are shown in Fig.~\ref{fig:case-study-A1-bode}, and the simulation results in Fig.~\ref{fig:case-study-A1}. It can be seen that the DVPP behaves according to the desired response. More specifically, the DC voltage and AC frequency responses well-match the re-aggregated desired responses using causalized $T_i$ or $T_i^{-1}$, i.e., $T_{{\mathrm{des}}}^{{\mathrm{vdcf}}-1} \left(\sum\nolimits_i T_i\right)^{-1}$, both in Fig.~\ref{fig:case-study-A1}(a) and (b). This suggests that all the converter local control strategies satisfactorily achieve their desired control behaviors.

In Fig.~\ref{fig:case-study-A1}(a), where the DC grid-supporting implementation is used, the measured response slightly deviates from (more precisely, better than) the original desired response. This deviation is due to the effect of the low-pass filters that are used to causalize $T_i$. In particular, the low-pass filters in $T_i$ produce a lead compensation in the re-aggregated frequency dynamics $T_{\mathrm{des}}^{\mathrm{vdcf}-1} \left( \sum\nolimits_i T_i \right)^{-1}$, as indicated by the Bode diagram in Fig.~\ref{fig:case-study-A1-bode}(a), introducing a dynamic damping effect and improving the dynamic performance. This implies that the basic inertial and droop response is not necessarily optimal and that we should have included a power oscillation damping service in the desired specification; see \cite{haberla2024optimal} for a systematic approach for choosing an optimal $T_{\mathrm{des}}^{\mathrm{fp}}$. Given this better damping (and easier power limiting), the DC grid-supporting implementation is recommended and used in the subsequent case studies.

\subsubsection{Case Study A2: Different Coordination Between DERs}

This case study aims to test the robustness of the control design under different disaggregation settings. First, we specify different time constants for the SC's inertial response: $T_{\mathrm{sc}} = -\frac{\tau_{\mathrm{sc}} s}{0.01s + 1}$, $\tau_{\mathrm{sc}} \in \{0.0, 0.1, 0.2, 0.3, 0.4\}$. Then, $T_{\mathrm{bess\&pv}}$ is disaggregated as in case study A1 such that the collective power response remains unchanged. The simulation results are shown in Fig.~\ref{fig:case-study-A2}(a), where the SC contributes different amounts of inertial power depending on the time constant. When the SC contributes less inertial power, the BESS delivers more inertial power. Moreover, we enable power limitations of $0.3$, $0.3$, and $0.7$\,pu in this case study for the SC, the BESS, and the PV, respectively. For small $\tau_{\mathrm{sc}} \in \{0, 0.1, 0.2\}$ assigned to the SC, the power response of the BESS reaches the power limit. As a result, the frequency response cannot well-match the desired, showing a faster rate of change of frequency (RoCoF). This suggests that the coordination of DERs should consider not only their dynamic characteristics but also their static capabilities/limitations. In Fig.~\ref{fig:case-study-A2}(b), we fixe $T_{\mathrm{sc}} = -\frac{0.3 s}{0.01s + 1}$ while specifying different static gains for the PV's participation factor: $T_{\mathrm{pv}} = \tfrac{\eta_{\mathrm{pv}}}{0.5s + 1} T_{\mathrm{bess\&pv}} $, $\eta_{\mathrm{pv}} \in \{0.6, 0.7, 0.8, 0.9, 1.0\}$, $T_{\mathrm{bess}} = \tfrac{0.5s + 1 - \eta_{\mathrm{pv}}}{0.5s + 1} T_{\mathrm{bess\&pv}}\frac{1}{(0.01s + 1)^2}$. It can be seen that the PV provides increasing static power as the static gain grows. The collective responses comply with the desired since the power limitations are not triggered.

\begin{figure}
  \begin{center}\vspace{-2mm}
  \includegraphics{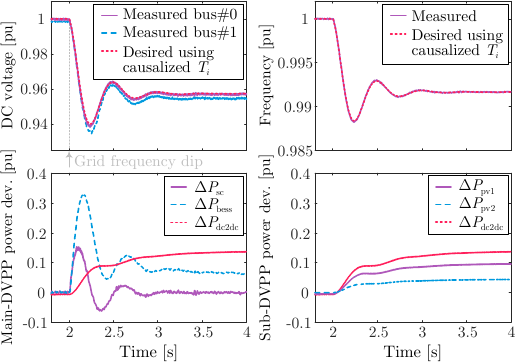}\vspace{-1mm}
  \caption{Simulation results of case study A3: the multi-DC-bus DVPP in Fig.~\ref{fig:case-study-system}(b) under a grid frequency dip.}
  \label{fig:case-study-A3}\vspace{-6mm}
  \end{center}
\end{figure}

\subsubsection{Case Study A3: Multi-DC-Bus DVPP Performance}

This case study tests the performance of a multi-DC-bus DVPP. The original PV is replaced with two small PVs (with $70$\% and $30$\% of the original PV capacity, respectively), as shown in Fig.~\ref{fig:case-study-system}(b), and we choose $T_{\mathrm{sc}}$, $T_{\mathrm{pv}}$, and $T_{\mathrm{bess}}$ the same as case study A1, whereas $T_{{\mathrm{des}}}^{{\mathrm{vdcvdc}}} = 1$, $T_{\mathrm{pv1}} = 0.7 \times T_{\mathrm{pv}}$, and $T_{\mathrm{pv2}} = 0.3 \times T_{\mathrm{pv}}$, all in per-unit. For the DC-coupled DC-output module, $T_{\mathrm{pv}}$, $T_{\mathrm{pv1}}$, and $T_{\mathrm{pv2}}$ are used to control the central, PV1's, and PV2' DC-DC converters, respectively. The simulation results are shown in Fig.~\ref{fig:case-study-A3}. We can see that the response of the entire DVPP matches the desired response, where the two PVs contribute in proportion $7$:$3$, as designed. Since the sub-DVPP delivers a little more power to compensate for power losses, the DC bus \#1's voltage is lower than the DC bus \#0's voltage in the post-disturbance steady state.

\subsubsection{Case Study A4: AC/DC Hybrid DVPP Performance}

\begin{figure}
  \begin{center}\vspace{-2mm}
  \includegraphics{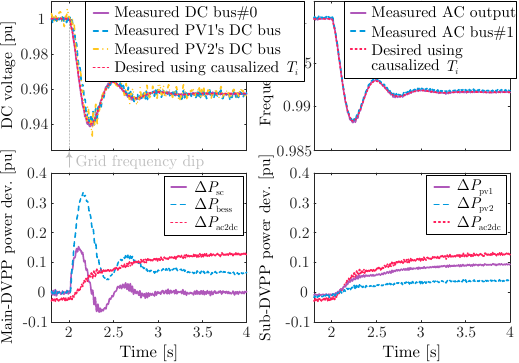}\vspace{-1mm}
  \caption{Simulation results of case study A4: the AC/DC hybrid DVPP in Fig.~\ref{fig:case-study-system}(c) under a grid frequency dip.}
  \label{fig:case-study-A4}\vspace{-6mm}
  \end{center}
\end{figure}

This case study tests the performance of a modular AC/DC hybrid DVPP. Instead of the DC-coupled DC-output module in case study A3, this case study introduces an AC-coupled DC-output module, with two small PVs, as illustrated in Fig.~\ref{fig:case-study-system}(c). Thus, this hybrid DVPP can represent a grid-connected AC/DC hybrid microgrid. The PV capacities and the disaggregation of the desired response are similar to case study A3. More specifically, $T_{\mathrm{pv1}} = 0.7 \times T_{\mathrm{pv}}$ and $T_{\mathrm{pv2}} = 0.3 \times T_{\mathrm{pv}}$ are used to control two PVs' DC-DC converters, respectively. Moreover, both the PVs' DC-AC inverters and the central AC-DC rectifier use $T_{\mathrm{des}}^{\mathrm{vdcf}{-1}}$ for AC-DC matching frequency-forming control and $T_{\mathrm{des}}^{\mathrm{vq}}$ for voltage-forming control. The simulation results are shown in Fig.~\ref{fig:case-study-A4}. It can be seen that the response of the entire DVPP satisfies the desired response, where the two PVs contribute power support in proportion $7$:$3$, as designed. Furthermore, due to the AC-DC matching relationship across the inverters and the rectifier, the DC bus~\#0 and the DC buses of both PVs' inverters share (almost) the same DC voltage response, and also the AC bus \#1 and the output-terminal AC bus share (almost) the same frequency response.

\subsection{Case Study B: Connection to the IEEE 13-Bus System}

\subsubsection{Case Study B1: Load Disturbance}

\begin{figure}
  \begin{center}
  \includegraphics{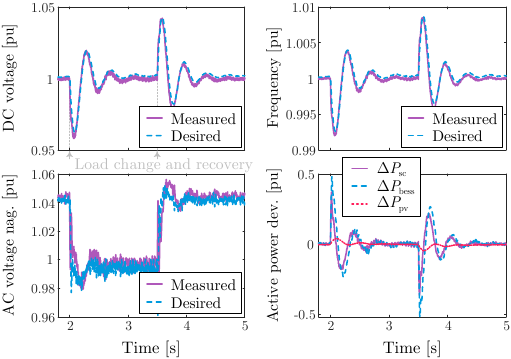}\vspace{-3mm}
  \caption{Simulation results of case study B1: the DVPP connected to the IEEE 13-bus system in Fig.~\ref{fig:case-study-system}(d) under load change and recovery.}
  \label{fig:case-study-B1}\vspace{-3mm}
  \end{center}
\end{figure}

This case study aims to validate the DVPP performance in a more realistic grid context. As shown in Fig.~\ref{fig:case-study-system}(d), a basic DVPP is connected through a transmission line to bus 680 of the IEEE 13-bus distribution system \cite{ieee13node}. A load disturbance of $5$ MW is imposed on bus 680. In the simulation result of Fig.~\ref{fig:case-study-B1}, the load disturbance occurs at $2$\,s and disappears at $3.5$\,s. We can see that the measured response to the disturbance matches the desired response well in terms of the DC voltage, the AC frequency, and the AC voltage magnitude. Moreover, since the frequency recovers to the nominal due to the presence of the main grid (modeled in an infinite bus), the DVPP only contributes an inertial response in terms of active power. Accordingly, the main contributors are the SC and the BESS.

\subsubsection{Case Study B2: Short-Circuit Fault}

\begin{figure}
  \begin{center}
  \includegraphics{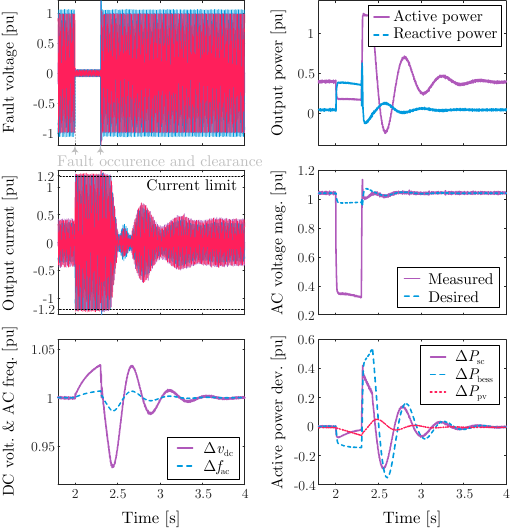}\vspace{-3mm}
  \caption{Simulation results of case study B2: the DVPP connected to the IEEE 13-bus system in Fig.~\ref{fig:case-study-system}(d) under a grid short-circuit fault.}
  \label{fig:case-study-B2}\vspace{-5mm}
  \end{center}
\end{figure}

This case study tests the fault ride-through performance of the DVPP in Fig.~\ref{fig:case-study-system}(d). A short-circuit fault is imposed on bus 680 for a duration of $0.3$\,s. To limit the fault current, the current limiter in the inverter control (as illustrated in Fig.~\ref{fig:control-block}) is enabled, and an additional feedback loop of current saturation information is activated during the current saturation \cite{he2024cross}. This is to maintain the grid-forming frequency/angle control while enforcing the current magnitude limiting for fault ride-through; see our recent work \cite{he2024cross} for further details on fault current limiting. The response of the DVPP under the grid fault is shown in Fig.~\ref{fig:case-study-B2}. The current magnitude is limited at a pre-specified level of $1.2$\,pu, and consequently, the AC voltage response deviates largely from the nominally desired response. The reactive power increases to respond to the voltage dip. Moreover, since the AC GFM and AC-DC matching functionalities are maintained during the grid fault \cite{he2024cross}, the DC voltage and the AC frequency are still under regulation even in the case of current saturation. Therefore, the DERs reduce their power outputs by utilizing the droop regulation characteristics when the DC voltage rises. After grid fault recovery and the current exits from saturation, the DVPP recovers to normal operation, matching the nominally desired response. The results of this case study validate the resilience of the proposed DVPP control under grid faults and demonstrate its compatibility with existing inner control and protection loops.

\vspace{-2mm}
\section{Conclusion}
\label{sec:conclusion}

This study presents a modular DVPP design as a flexible and scalable solution to the coordinated GFM control of heterogeneous DERs. Modular DVPPs can be topologically based on AC coupling, DC coupling, or hybrid coupling. Particularly, for a DC-coupled AC-output DVPP, the proposed control design allows for a desired GFM response, AC-DC matching behavior, and coordinated power contribution between heterogeneous DERs, beyond the localized self-contained functionality of conventional microgrids. Based on the four basic DVPP modules, we present a systematic top-down approach for constructing and controlling modular DVPPs and more generally modular AC/DC hybrid power grids.

There are several critical directions for further research. One is to explore top-down paradigms to design, shape, and manage the dynamic behavior of AC/DC hybrid power grids, in contrast to traditional bottom-up paradigms that rely on modeling, analysis, and design from device-level individual components to the grid-level whole system. Another direction is the development of standardized protocols or new-generation grid codes to promote the widespread adoption and interoperability of such top-down design paradigms.

\appendix

This appendix shows the control objectives and control designs of the other three basic DVPP modules that are not detailed in the main text. The aggregation conditions are focused, as the disaggregation and implementation are similar to the previous DC-coupled AC-output module.

\textit{1) AC-Coupled DC-Output DVPP Module:} We specify the desired power response $\Delta p$ of the central rectifier to an output-terminal DC voltage variation $\Delta (v_{{\mathrm{dc}}}^2)$ as
\begin{equation}
    \label{eq:rectifier-desired}
    \Delta p(s) = T^{\mathrm{pvdc}}_{\mathrm{des}}(s) \Delta (v_{{\mathrm{dc}}}^2)(s).
\end{equation}
Similar to \eqref{eq:ac-dc-matching}, we specify a desired AC-DC matching relationship across the rectifier as
\begin{equation}
\label{eq:rectifier-matching}
    \Delta (v_{{\mathrm{dc}}}^2)(s) = T_{{\mathrm{des}}}^{{\mathrm{vdcf}}}(s) \Delta {f}(s),
\end{equation}
where $\Delta f$ denotes the frequency variation of the AC-coupling bus. Moreover, the frequency response of each inverter connecting the DER is specified as
\begin{equation}
\label{eq:inverter}
    \Delta f_i(s) = T_i(s) \Delta p_i(s),
\end{equation}
with the frequency-forming variable $\Delta f_i$ and the measured power change $\Delta p_i$. Based on this, the coherent frequency at the AC-coupling bus can be derived as \cite{haberle2024gridforming}
\begin{equation}
\label{eq:coherent}
    \Delta f_{\mathrm{sync}}(s) = \left( \textstyle\sum\nolimits_i T_i(s)^{-1} \right)^{-1} \textstyle\sum\nolimits_i \Delta p_i(s).
\end{equation}
From \eqref{eq:rectifier-desired}, \eqref{eq:rectifier-matching}, \eqref{eq:coherent}, the frequency coherency $\Delta {f} \approx \Delta f_{\mathrm{sync}}$, and the power conservation $\Delta p = \sum\nolimits_i \Delta p_i$, the aggregation condition for the AC-coupled DC-output DVPP is derived as
\begin{equation}
    \label{eq:ac-dc-aggregation-condition}
    \textstyle\sum\nolimits_i T_i(s)^{-1} =  T_\mathrm{des}^{\mathrm{vdcf}}(s) T_{{\mathrm{des}}}^{{\mathrm{pvdc}}}(s).
\end{equation}

\textit{2) AC-Coupled AC-Output DVPP Module:} Similar to \eqref{eq:ac-gfm} as for a DC-coupled AC output module, we specify the desired frequency response $\Delta f$ as
\begin{equation}
    \label{eq:ac-desired}
    \Delta f(s) = T_\mathrm{des}^{\mathrm{fp}}(s) \Delta p(s),
\end{equation}
where $\Delta p$ is the active power change at the output terminal. The frequency response of each inverter can be specified as in \eqref{eq:inverter}. We can immediately obtain the aggregation condition from the coherent frequency dynamics in \eqref{eq:coherent} as \cite{haberle2024gridforming}
\begin{equation}
\label{eq:ac-ac-aggregation-condition}
    \left( \textstyle\sum\nolimits_i T_i(s)^{-1} \right)^{-1} = T_{\mathrm{des}}^{\mathrm{fp}}(s).  
\end{equation}

\textit{3) DC-Coupled DC-Output DVPP Module:} We specify the desired power response $\Delta p$ of the central DC-DC converter to an output-terminal DC voltage variation $\Delta (v_{{\mathrm{dc}}}^2)$ as
\begin{equation}
    \label{eq:dc2dc-desired}
    \Delta p(s) = T^{\mathrm{pvdc}}_{\mathrm{des}}(s) \Delta (v_{{\mathrm{dc}}}^2)(s).
\end{equation}
Likewise, we desire a DC-DC matching relationship between the input voltage $\Delta (v_{{\mathrm{dc}},1}^2)$ and the output voltage $\Delta (v_{{\mathrm{dc}}}^2)$ as
\begin{equation}
\label{eq:dc2dc-matching}
    \Delta (v_{{\mathrm{dc}},1}^2)(s) = T_{{\mathrm{des}}}^{{\mathrm{vdcvdc}}}(s) \Delta (v_{{\mathrm{dc}}}^2)(s).
\end{equation}
Furthermore, as in \eqref{eq:DER-dynamics}, the power response of each DC-DC converter connecting the DER is represented as
\begin{equation}
\label{eq:dc2dcconverter}
    \Delta p_i(s) = T_i(s) \Delta (v_{{\mathrm{dc}},1}^2)(s).
\end{equation}
From \eqref{eq:dc2dc-desired} to \eqref{eq:dc2dcconverter} and the power conservation relationship $\Delta p = \sum\nolimits_i\Delta p_i(s)$, the aggregation condition for a DC-coupled DC-output DVPP module is derived as
\begin{equation}
    \label{eq:dc-dc-aggregation-condition}
    \textstyle\sum\nolimits_i T_i(s) = T_{{\mathrm{des}}}^{{\mathrm{vdcvdc}}}(s)^{-1} T^{\mathrm{pvdc}}_{\mathrm{des}}(s).
\end{equation}


The parameters used in the case studies are listed in Table~\ref{tab:system-parameters}.

\begin{table}
\scriptsize
\centering
\caption{Parameters in Simulation Validations}\vspace{-5mm}
\begin{tabular}[t]{lll}
\arrayrulecolor{black}
\hline \hline
Symbol                 & Description       & Value               \\
\hline
\hline
$s_N$       & DVPP nominal capacity  & $1$\,MVA \\
$\omega_0$  & Fundamental frequency & $120\pi$\,rad/s\\
$v_N$       & AC nominal voltage (ph-ph, rms) & $400$\,V\\
$v^{\star}_{\mathrm{dc,0}}$   & DC bus \#0 nominal voltage & $800$\,V\\
$v^{\star}_{\mathrm{dc,1}}$   & DC bus \#1 nominal voltage & $480$\,V\\
$\pha z_{\mathrm{g}}$ & Grid impedance  & $0.08 + j0.4$\,pu\\
$\pha z_{\mathrm{l}}$ & Line impedance  & $0.07 + j0.2$\,pu\\
$\pha z_{\mathrm{t}}$ & Transformer impedance & $0.002 + j0.06$\,pu\\
$l_\mathrm{f}$      & Filter inductance & $0.10$\,pu\\
$r_\mathrm{f}$      & Filter inductance & $0.0005$\,pu\\
$c_\mathrm{f}$      & Filter inductance & $0.08$\,pu\\
$c_\mathrm{dc,0}$     & DC capacitor on bus \#0 & $80000\,\mu$F\\
$c_\mathrm{dc,1}$     & DC capacitor on bus \#1 & $10000\,\mu$F\\
$c_\mathrm{sc}$     & Supercapacitor (SC) & $20000$\,F\\
$c_\mathrm{bc}$     & Boost converter's capacitor & $5000\,\mu$F\\
$l_\mathrm{bc}$     & Boost converter's inductor & $0.4$\,mH\\
$r_\mathrm{bc}$     & Boost converter's resistor & $1$\,m$\Omega$\\
$r_{\mathrm{l}}$     & DC cable resistor & $1$\,m$\Omega$\\
$l_{\mathrm{l}}$     & AC cable inductor & $0.6$\,mH\\
$p^{\star}_{\mathrm{sc}}$ & SC setpoint & $0.0$\,pu \\
$p^{\star}_{\mathrm{bess}}$ & BESS setpoint & $0.0$\,pu \\
$p^{\star}_{\mathrm{pv}}$ & PV setpoint & $0.5$\,pu \\
$p_{\mathrm{dc,load}}$ & Constant-power load on bus \# 0 & $0.1$\,pu \\
$q^{\star}$ & Inverter reactive power setpoint & $0.0$\,pu \\
$v^{\star}$ & Inverter voltage setpoint & $1.05$\,pu \\
$\pha {z}_\mathrm{v}$      & Virtual impedance & $j0.1$\,pu \\
$I_{\lim}$ & Current limit & $1.2$\,pu \\
$r_{\mathrm{f}}$ & Fault resistance at bus 680 & $0.01\,\Omega$ \\
$\Delta p_{\mathrm{load}}$ & Load disturbance at bus 680 & $5$\,MW \\
$f_{\mathrm{sw}}$ & Switching frequency & $5$\,kHz \\
$T_{\mathrm{s}}$ & Simulation step size & $5$\,$\mu$s \\
\hline \hline
\end{tabular}\vspace{-4mm}
\label{tab:system-parameters}
\end{table}

\bibliographystyle{IEEEtran}
\bibliography{IEEEabrv,Bibliography}

\end{document}